\documentclass[aps,prd,preprintnumbers,nofootinbib,twocolumn]{revtex4}
\usepackage{bm}
\usepackage{latexsym}
\usepackage{dcolumn}
\usepackage{amsmath,amsfonts,amssymb}
\usepackage{graphicx,epsfig}
\usepackage{psfrag}
\usepackage{amsthm}

\def\be {\begin{equation}}
\def\ee {\end{equation}}
\def\bea {\begin{eqnarray}}
\def\eea {\end{eqnarray}}
\def\bc {\begin{center}}
\def\ec {\end{center}}
\def\bfg {\begin{figure}}
\def\efg {\end{figure}}
\def\bi {\begin{itemize}}
\def\ei {\end{itemize}}
\def\nn {\nonumber}

\def\la {\label}
\def\le {\left}
\def\ri {\right}

\def\fr {\frac}

\def\no {\noindent}

\def\a  {\alpha}

\def\D  {\Delta}
\def\e  {\epsilon}

\def\beq{\begin{equation}}
\def\eeq{\end{equation}}
\def\br{\begin{eqnarray}}
\def\er{\end{eqnarray}}
\newcommand{\eel}[1] {\label{#1}\end{equation}}

\newcommand{\bdm}{\begin{displaymath}}
\newcommand{\edm}{\end{displaymath}}

\begin{document}
\title{Discreteness of Space from the Generalized Uncertainty Principle}

\author{Ahmed Farag Ali $^1$} \email[email: ]{ahmed.ali@uleth.ca}
\author{Saurya Das $^1$} \email[email: ]{saurya.das@uleth.ca}
\author{Elias C. Vagenas $^2$} \email[email: ]{evagenas@academyofathens.gr}

\affiliation{$^1$~Dept. of Physics,
University of Lethbridge, 4401 University Drive,
Lethbridge, Alberta, Canada T1K 3M4 \\}

\affiliation{$^2$~Research Center for Astronomy \& Applied Mathematics,\\
Academy of Athens, \\
Soranou Efessiou 4,
GR-11527, Athens, Greece
}

\begin{abstract}
%
Various approaches to Quantum Gravity (such as String Theory and
Doubly Special Relativity), as well as black hole physics
predict a minimum measurable
length, or a maximum observable momentum, and related modifications
of the Heisenberg Uncertainty Principle to a so-called
Generalized Uncertainty Principle (GUP). We propose a GUP
consistent with String Theory, Doubly Special Relativity and black hole
physics, and show that this modifies all quantum mechanical Hamiltonians. When
applied to an elementary particle, it implies that the space which confines it
must be quantized. This suggests that space itself
is discrete, and that all measurable lengths are quantized
in units of a fundamental length (which can be the
Planck length). On the one hand, this signals
the breakdown of the spacetime continuum picture near that
scale, and on the other hand, it can predict an upper bound
on the quantum gravity parameter in the GUP, from current
observations. Furthermore, such fundamental discreteness of space may
have observable consequences at length scales much larger than the
Planck scale.
\end{abstract}

\maketitle


An intriguing prediction of various theories of quantum gravity
(such as String Theory) and black hole physics is the existence of
a minimum measurable length. This has given rise to the so-called
Generalized Uncertainty Principle, or GUP, or equivalently,
modified commutation relations between position coordinates and
momenta \cite{guppapers}. The recently proposed {\it Doubly
Special Relativity} (or DSR) theories on the other hand (which
predict maximum observable momenta), also suggest a similar
modification of commutators \cite{sm,cg}. The commutators which
are consistent with String Theory, Black Holes Physics, DSR, {\it
and} which ensure $[x_i,x_j]=0=[p_i,p_j]$ (via the Jacobi
identity) have the following form \cite{asv1}
\footnote{
The results of this article
do not depend on this particular form of GUP chosen,
and continue to hold for a a large class of variants,
so long as an ${\cal O}(\a)$ term is present in the right hand side
of Eq.(\ref{comm01}).
}
\bea
[x_i, p_j] = i \hbar\hspace{-0.5ex} \left[  \delta_{ij}\hspace{-0.5ex}
- \hspace{-0.5ex} \alpha\hspace{-0.5ex}  \le( p \delta_{ij} +
\frac{p_i p_j}{p} \ri)
+ \alpha^2 \hspace{-0.5ex}
\le( p^2 \delta_{ij}  + 3 p_{i} p_{j} \ri) \hspace{-0.5ex} \ri]
\label{comm01}
\eea
where
$p^{2} = \sum\limits_{j=1}^{3}p_{j}p_{j} $, $\alpha = {\alpha_0}/{M_{Pl}c}
= {\alpha_0 \ell_{Pl}}/{\hbar},$
$M_{Pl}=$ Planck mass, $\ell_{Pl}\approx 10^{-35}~m=$ Planck length,
and $M_{Pl} c^2=$ Planck energy $\approx 10^{19}~GeV$.
Eq.(\ref{comm01}) gives, in $1$-dimension, to ${\cal O}(\a^2)$
\bea
&& \Delta x \D p \geq \frac{\hbar}{2}
\le[
1 - 2\a <p> + 4\a^2 <p^2>
\ri] ~ \nn \\
&\geq& \hspace{-1ex}
\frac{\hbar}{2} \hspace{-1ex}
\le[ 1\hspace{-0.5ex}  +\hspace{-0.5ex}  \le(\hspace{-0.5ex}  \frac{\a}{\sqrt{\langle p^2 \rangle}} +4\a^2 \hspace{-0.5ex} \ri)
\hspace{-0.5ex}  \D p^2 \hspace{-0.5ex}
+\hspace{-0.5ex}  4\a^2 \langle p \rangle^2 \hspace{-0.5ex}
- \hspace{-0.5ex}  2\a \sqrt{\langle p^2 \rangle}
\ri]\hspace{-1ex} . \label{dxdp1}
\eea
Commutators and inequalities similar to (\ref{comm01}) and
(\ref{dxdp1}) were proposed and derived respectively in
\cite{kmm,kempf,brau,dv}. These in turn imply a minimum measurable
length {\it and} a maximum measurable momentum (to the best of our
knowledge, (\ref{comm01}) and (\ref{dxdp1}) are the only forms
which imply both)

\bea
\D x &\geq& (\D x)_{min}  \approx \alpha_0\ell_{Pl} \la{dxmin} \\
\D p &\leq& (\D p)_{max} \approx \frac{M_{Pl}c}{\a_0}~. \la{dpmax}
\eea
\par\noindent
Next, defining \cite{asv1}
\bea x_i = x_{0i}~,~~
p_i = p_{0i} \le( 1 - \a p_0 + 2\a^2 p_0^2 \ri)~, \la{mom1}
\eea
%
with $x_{0i}, p_{0j}$
satisfying the canonical commutation relations
%
$ [x_{0i}, p_{0j}] = i \hbar~\delta_{ij}, $
%
it can be shown that Eq.(\ref{comm01}) is satisfied. Here,
$p_{0i}$ can be interpreted as the momentum at low energies
(having the standard representation in position space, i.e. $p_{0i} = -i
\hbar \partial/\partial{x_{0i}}$), $p_{i}$ as that at higher energies, and
$p_0$ as the magnitude of the $p_{0i}$ vector, i.e. $p_{0}^{2} =
\sum\limits_{j=1}^{3}p_{0j}p_{0j}$.
%
%
%
%
It is normally assumed that the dimensionless parameter
$\a_0$ is of the order of unity, in which case
the $\a$ dependent terms are important only when
energies (momenta) are comparable to the Planck energy (momentum),
and lengths are comparable to the Planck length.
However, we do not impose this
condition {\it a priori}, and note that this may signal the
existence of a new physical length scale of the order of
$\a\hbar=\a_o\ell_{Pl}$. Evidently, such an intermediate length
scale cannot exceed the electroweak length scale $\sim 10^{17}~\ell_{Pl}$ (as
otherwise it would have been observed). This implies $\a_0 \leq 10^{17}$.

Using (\ref{mom1}), a Hamiltonian of the form
\bea H &=& \fr{p^2}{2m} + V(\vec r)
\eea
can be written as
\bea
H&=&H_0 + H_1 + {\cal O}( \a^2) ~, \\
%
%
\mbox{where}~H_0 &=& \fr{p_0^2}{2m} + V(\vec r)   \\
\mbox{and}~ H_1 &=& -\fr{\a}{m}~p_0^3~. 
\eea
Thus, we see that {\it any} system with a
well defined quantum (or even classical) Hamiltonian $H_0$, is perturbed
by $H_1$, defined above, near the Planck scale. In other words,
Quantum Gravity effects are in some sense universal!
The relativistic Dirac equation is modified in a similar way, and
is expected to give rise to the main result of this paper \cite{asv2}, which is
unaffected by the inclusion of the
${\cal O}(\a^2)$ terms in $H_1$ as well.
Phenomenological implications of the GUP in diverse quantum systems
have been studied (for example see \cite{dv2} and references therein).

In this article, we apply the above formalism to
a single particle in a box of length $L$
(with boundaries at $x=0$ and $x=L$), and show that the box length
must be quantized.
Since this particle can be considered as {\it test particle}
to measure the dimension of the box, this suggests that space
itself is quantized, as are all observable lengths.
The wavefunction of the particle
satisfies the following GUP corrected Schr\"odinger equation
inside the box, where
$V(\vec r)=0$ (outside, $V=\infty$ and $\psi=0$)
\be
H \psi = E \psi~,
\ee
or equivalently ($d^n \equiv d^n/dx^n$),
\bea
&& d^2\psi + k^2 \psi + 2i\alpha \hbar d^3 \psi = 0~,
\la{seqn}
%
%
%
\eea
where
$ k=\sqrt{2mE/\hbar^2}$.
A trial solution of the form $\psi=e^{mx}$ yields
\be
m^2 + k^2 + 2i\a \hbar m^3 =0~,
\ee
with the following solution set to leading order in $\a$: $m=\{ik',-ik'',i/2\a\hbar\}$,
where $k'=k(1+k\a\hbar)$ and $k''=k(1-k\a\hbar)$.
Thus, the general wavefunction to leading order in $\ell_{Pl}$ and $\a$ is of the form
\bea
\psi &=& Ae^{ik'x} + B e^{-ik''x} + C e^{ix/2\a\hbar}  . \la{barrierpsi1}
\la{wavefn1}
\eea
As is well known, the first two terms (with $k'=k''=k$) and the boundary conditions
$\psi=0$ at $x=0$, $L$ give rise to the standard quantization of energy for a particle
in a box, namely $E_n=n^2\pi^2\hbar^2/2mL^2$.
However, note the appearance of a new oscillatory term here, with characteristic wavelength
$4\pi \a\hbar$ and momentum $1/4\alpha = M_{Pl}c/4\alpha_0$ (which is Planckian for $\alpha_0 = {\cal O}(1)$).
This results in the new quantization mentioned above.
Also, as this term should drop out in the $\a\rightarrow 0$ limit, one must have
$\lim_{\a\rightarrow 0} |C|=0$.
We absorb any phase of $A$ in $\psi$,
such that $A$ is real. The boundary condition
\be
\psi(0) = 0
\ee
implies
\be
A + B + C = 0~.
\la{bc1}
\ee
Substituting for $B$ in Eq.(\ref{wavefn1}), we get
\bea
\psi &=& 2i A \sin(kx) + C \le[ - e^{-ikx} + e^{ix/2\a\hbar}\ri] \nn \\
&-& \a\hbar k^2x \le[
i~C e^{-ikx} + 2 A \sin(kx)
\ri]~.
\eea
The remaining boundary condition
\be
\psi(L) = 0
\ee
yields
\bea
2iA \sin(kL) &=& \left|C\right| \le[
e^{-i(kL+\theta_C)}
-e^{i(L/2\a\hbar - \theta_C)}
\ri] \nn \\
&+& \hspace{-0.5ex}\a\hbar k^2 L \hspace{-0.5ex}
\le[
i\le|C\ri| e^{-i(kL+\theta_C)}\hspace{-0.5ex} +\hspace{-0.5ex} 2A\sin(kL)
\ri]
\hspace{4ex}
\label{master}
\eea
where $C = |C| \exp(-i\theta_C)$.
Note that both sides of the above equation
vanish in the limit $\a \rightarrow 0$, when $kL=n\pi~(n \in \mathbb{Z})$
and $C=0$. Thus, when $\a \neq 0$, we must have
$kL = n\pi + \epsilon$, where $\epsilon \in \mathbb{R}$ (such that energy eigenvalues
$E_n$ remain positive), and
$\lim_{\a\rightarrow 0}\epsilon =0 .$ This, along with the previously discussed smallness
of $|C|$ ensures that the second line in Eq.(\ref{master}) above falls off faster than
${\cal O}(\a)$, and hence can be dropped.
Next, equating the real parts of the remaining terms of
Eq.(\ref{master}) (remembering that $A\in\mathbb{R}$), we get
\bea
\cos\le(\hspace{-0.5ex} \frac{L}{2\a\hbar} - \theta_C \hspace{-0.5ex}\ri)
\hspace{-0.5ex}=\hspace{-0.5ex}cos\le( kL + \theta_C \ri)
\hspace{-0.5ex}=\hspace{-0.5ex}\cos\le( n\pi + \theta_C + \epsilon \ri)~,\hspace{1.5ex}
\eea
which implies, to leading order, the following two series of solutions
\bea
\frac{L}{2\alpha\hbar} &=& \frac{L}{2\a_0 \ell_{Pl}} = n\pi + 2q \pi +2\theta_C
\equiv p\pi + 2\theta_C \label{soln1}\\
\frac{L}{2\alpha\hbar} &=& \frac{L}{2\a_0 \ell_{Pl}} = -n\pi + 2q\pi
\equiv p \pi  ~, \label{soln2}\\
p &\equiv& 2q \pm n \in \mathbb{N}. \nn
\eea

These show that there cannot even be a single particle in the
box, unless its length is quantized as above. For other lengths,
there is no way to probe or measure the box, even if it exists.
Hence, effectively all measurable lengths are
quantized in units of $\a_0\ell_{Pl}$! We interpret this as
space essentially having a discrete nature.
Consistency with Eq.(\ref{dxmin}) requires
$p$ to run from $1$ in the second case.
The minimum length is $\approx \a_0 \ell_{Pl}$ in each case. Once
again, if $\a_0 \approx 1$,
this fundamental unit is the Planck length. However, current experiments
do not rule out discreteness smaller than about a thousandth of a
Fermi, thus predicting the previously mentioned bound on
$\a_0$
\footnote{
Equating the imaginary parts of (\ref{master}) yields the auxiliary
condition: $\epsilon=-\left|C\right|\sin(\theta_C)/A$ and $\e=0$, for solutions
(\ref{soln1}) and (\ref{soln2}) respectively.}.
Note that similar quantization of length was shown in the context
of Loop Quantum Gravity in \cite{thiemann}, albeit following a
much more involved analysis, and perhaps under a stronger set
of starting assumptions. In general however, we expect our result
to emerge from any correct theory of Quantum Gravity.
It will be interesting to see whether our
result can be generalized to the quantization of areas and
volumes, and also to study its possible phenomenological implications.
Furthermore, it is plausible that if space has fundamentally a ``grainy'' structure,
the effects may be felt well beyond the Planck scale, e.g.
at around $10^{-4}~fm$, the length scale to be probed at the
Large Hadron Collider (similar to Brownian motion observed at scales
in excess of $10^5$ times the atomic scale.).
We hope to study such effects and report elsewhere.

\no {\bf Acknowledgment}
We thank A. Dasgupta, S. Hossenfelder and L. Smolin for interesting discussions.
This work was supported in part by the Natural
Sciences and Engineering Research Council of Canada and by the
Perimeter Institute for Theoretical Physics.




\begin{thebibliography}{99}

\bibitem{guppapers} D. Amati, M. Ciafaloni, G. Veneziano,
Phys. Lett. B {\bf 216} (1989) 41;
M.~Maggiore,
 Phys.\ Lett.\  B {\bf 304} (1993) 65  [arXiv:hep-th/9301067];
M.~Maggiore,
  Phys.\ Rev.\  D {\bf 49} (1994) 5182 [arXiv:hep-th/9305163];
M.~Maggiore,
  Phys.\ Lett.\  B {\bf 319} (1993) 83
  [arXiv:hep-th/9309034];
L.~J.~Garay, Int.\ J.\ Mod.\ Phys.\  A {\bf 10} (1995) 145 [arXiv:gr-qc/9403008];
%
F.~Scardigli,
  Phys.\ Lett.\  B {\bf 452} (1999) 39
  [arXiv:hep-th/9904025];
  %
  %
S.~Hossenfelder, M.~Bleicher, S.~Hofmann, J.~Ruppert, S.~Scherer and H.~Stoecker,
  Phys.\ Lett.\  B {\bf 575} (2003) 85
  [arXiv:hep-th/0305262];
  %
  %
C.~Bambi and F.~R.~Urban,
  Class.\ Quant.\ Grav.\  {\bf 25} (2008) 095006
  [arXiv:0709.1965 [gr-qc]].

\bibitem{sm} J.~Magueijo and L.~Smolin,
  Phys.\ Rev.\ Lett.\  {\bf 88} (2002) 190403
  [arXiv:hep-th/0112090];
J.~Magueijo and L.~Smolin,
  Phys.\ Rev.\  D {\bf 71} (2005) 026010
  [arXiv:hep-th/0401087].
.

\bibitem{cg}
J. L. Cortes, J. Gamboa, Phys. Rev. D {\bf 71} (2005) 065015
[arXiv:hep-th/0405285];

\bibitem{asv1}
A.~F.~Ali, S.~Das, E.~C.~Vagenas, paper in preparation.


\bibitem{kmm} A. Kempf, G. Mangano, R. B. Mann, Phys. Rev. D {\bf
 52} (1995) 1108 [arXiv:hep-th/9412167].

\bibitem{kempf} A. Kempf, J.Phys. A  {\bf 30} (1997) 2093 [arXiv:hep-th/9604045].

\bibitem{brau} F. Brau, J. Phys. A {\bf  32} (1999) 7691 [arXiv:quant-ph/9905033].



\bibitem{dv} S. Das, E. C. Vagenas, Phys. Rev. Lett. {\bf 101} (2008) 221301 [arXiv:0810.5333 [hep-th]].

\bibitem{asv2}
A.~F.~Ali, S.~Das, E.~C.~Vagenas, paper in preparation.

\bibitem{dv2} S. Das and E. C. Vagenas,
Can.\ J.\ Phys.\  {\bf 87} (2009) 233 [arXiv:0901.1768 [hep-th]].

\bibitem{thiemann} T. Thiemann, J. Math. Phys. {\bf 39} (1998) 3372-3392 [arXiv:gr-qc/9606092].

\end{thebibliography}
\end{document}